**Spinodals and Domain Instabilities in Ionically-compensated Ferroelectric Films**

Sergei V. Kalinin[1]

Department of Materials Science and Engineering and Electrical Engineering and Computer Science, University of Tennessee, Knoxville, Tennessee 37996, USA

**Abstract**

A ferroelectric surface in contact with an ionic or molecular reservoir is intrinsically a ferroionic state in which polarization and compensating surface charge are thermodynamically coupled. Surface chemical compensation is known to control the magnitude and sign of ferroelectric polarization, yet the competing instability toward domain formation has usually been treated separately. Here we develop a static theory in which polarization, electrochemical surface charge, and finite-wave-vector domain formation are analyzed within a single thermodynamic framework for uniaxial polarization orientation. The central quantity is the differential chemical capacitance, which determines how strongly the surface charge can respond to a spatial modulation of polarization and therefore differs fundamentally from the equilibrium screening charge. We derive the homogeneous equation of state, the finite-wave-vector stability kernel, analytical domain-onset wavelengths and thickness scalings, and weakly nonlinear stripe and checkerboard solutions for both second- and first-order ferroelectrics. For a first-order transition, a finite-amplitude stripe state can become thermodynamically favorable before the homogeneous linear spinodal when the effective quartic coefficient is negative; this pre-spinodal window closes as the biased background polarization increases. Dense numerical phase maps confirm the analytical structure and show how chemical potential and film thickness reorganize weakly polar, monodomain, and polydomain states. Numerical analysis for $BaTiO_3$ realization illustrates the corresponding temperature, oxygen-pressure, and thickness scales and identifies experimentally testable regimes in which chemical screening stabilizes the homogeneous state against the columnar 180° instability considered here or domain formation becomes the preferred route for depolarization-energy reduction.

[1] sergei2@utk.edu

## I. Introduction

Ferroelectric polarization and surface chemistry are intrinsically coupled because the normal component of polarization generates bound surface charge. At an ideal metal this charge can be compensated electronically, whereas an exposed surface can exchange matter and charge with its environment through adsorption, dissociation, redox processes, vacancies, protons or hydroxyls, and other ionic or molecular species. The electrochemical populations of these species depend on the local surface potential, while that potential is itself generated by the polarization and the compensating charge. The natural thermodynamic loop is therefore polarization → surface potential → surface chemistry → screening charge → depolarization field → polarization. This reciprocity is the basis of what is now commonly described as electrochemical–ferroelectric or ferroionic coupling [1–5]. It implies that chemical potential, oxygen pressure, humidity, and other environmental variables can act as thermodynamic control parameters for ferroelectricity rather than merely as extrinsic perturbations. The existing studies have largely emphasized the homogeneous consequences of this coupling: chemical stabilization of one polarization orientation, suppression or displacement of the ferroelectric transition, modified switching, emergence of mixed electrochemical–ferroelectric states, and slow relaxation when the chemical and polar degrees of freedom evolve on different timescales [2–13]. Atomistic calculations show that water chemisorption and hydroxyl dipoles can couple directly to polarization on $BaTiO_3$ surfaces [14], and controlled-humidity PFM demonstrates that adsorbed water and associated ionic mobility modify domain-writing speed and wall kinetics [15].

It is important to note that the physical idea substantially predates the terminology. Parravano reported a connection between ferroelectric transitions and heterogeneous catalysis in 1952 [16]. Hooton and Merz showed in 1955 that opposite polar terminations of $BaTiO_3$ etch at different rates [17], while Känzig and Chynoweth established that surface space charge is an essential part of the electrostatic state of a ferroelectric [18,19]. Toyoda and Itakura subsequently demonstrated that the atmosphere and adsorption of polar molecules can alter polarization-reversal kinetics in $BaTiO_3$ [20]. Surface-science and catalysis studies in the 1970s–1990s then showed polarization-dependent adsorption and catalytic behavior on $KNbO_3$ and $LiNbO_3$ [21–24]. These observations already contained both directions of the modern feedback loop: polarization can alter surface chemistry, and surface/environmental state can alter ferroelectric behavior. What was largely absent before 2000 was a domain-resolved measurement of the screening state and a thermodynamic formalism in which a chemical reservoir supplied the compensating charge self-consistently. A detailed chronology and discussion of the pre-2000 literature are given in Supplementary Section S1.

Beginning around 2000, scanning probe and synchrotron measurements transformed this qualitative picture into a local and thermodynamic framework. Kalinin and Bonnell used electrostatic-force and Kelvin Probe Force Microscopy to show that the polarization charge of ambient $BaTiO_3$ surfaces can be completely compensated and that the measured domain potential is consistent with an adsorbate-mediated surface double layer [1]; subsequent work connected local polarization, screening, and surface reactivity [25,26]. In parallel, domain-selective photochemistry and surface-science measurements established polarization-dependent reaction and adsorption energetics [27–31]. The converse effect was demonstrated by Wang, Fong, Highland, Stephenson, and collaborators using *in situ* synchrotron scattering: changing oxygen partial pressure reversibly changed the polarization orientation of ultrathin $PbTiO_3$ [6], and the equilibrium Curie temperature and polarization state were mapped as a function of oxygen pressure [8]. Stephenson and Highland formulated the associated electrochemical boundary-condition theory by coupling Landau thermodynamics to ionic surface equilibria [2]. Morozovska, Eliseev, and Kalinin subsequently generalized this logic into the explicit ferroionic-state framework, in which surface ionic charge and polarization are mutually coupled state variables [4]. The broader screening and domain literature was synthesized by Kalinin, Kim, Fong, and Morozovska [5]. These developments established the homogeneous electrochemical thermodynamics on which the present work builds. The same period also established nanoscale 180° stripe domains by x-ray scattering [32], a first-principles critical-thickness problem for imperfectly screened ultrathin films [33], and polarization-directed multicomponent ferroelectric lithography [34].

However, these studies did not analyze the emergence of domain structures in electrochemically screened ferroelectrics. Here, incomplete chemical compensation can be relieved not only by reducing or reversing the homogeneous polarization, but also by forming domains. This possibility is fundamental because domain formation is itself the classical electrostatic response of a ferroelectric to imperfect screening [35–38]. A chemically open film therefore has at least three competing routes: change the homogeneous order parameter, change the surface chemical charge, or create a spatially modulated polarization pattern. The relevant chemical descriptor is not only the equilibrium surface charge, σ, but also its differential response to surface potential. We define this response as the differential chemical capacitance, $C_{\mathrm{chem}} = -\partial\sigma/\partial\phi_s$. A surface can carry nearly the amount of charge required to compensate the average polarization while having a small differential capacitance, for example when the relevant adsorbate population is saturated. Such a surface can stabilize the mean polarization yet screen a newly developing finite-wave-vector fluctuation poorly. This distinction makes chemical capacitance a natural organizing variable for the onset of domain formation. The thickness dependence of ferroic stripe patterns and the role of wall width provide an additional bridge between classical Kittel scaling and finite-size domain thermodynamics [39].

Here we formulate this competition for a uniaxial ferroelectric film with an ideal grounded bottom electrode and an electrochemically active upper surface. We first derive the homogeneous equation of state and its linearized chemical boundary condition. We then allow lateral polarization modulations and obtain an analytical finite-wave-vector stability kernel that contains the chemical capacitance explicitly. In the weak-screening limit the domain-onset wavelength scales as $h^{1/3}$, whereas developed sharp-wall domains recover a Kittel-like $h^{1/2}$ law. Second- and first-order ferroelectrics exhibit differing behaviors, when the former undergo a continuous finite-wave-vector instability, while in the latter a finite-amplitude stripe state can become stable before the homogeneous spinodal is reached only while the effective quartic coefficient $\beta + 10\gamma P_0^2$ remains negative; at larger biased $P_0$ the onset becomes continuous. A comparison of stripes and checkerboards shows that stripes are preferred in the scalar isotropic model both close to onset and in the developed-domain limit. Dense numerical calculations map the resulting temperature–chemical-potential–thickness phase structure. Finally, we apply the framework to $BaTiO_3$ using an eighth-order bulk thermodynamic potential and literature gradient/dielectric parameters, while treating the oxygen surface chemistry as an explicitly declared effective boundary model. The key new result is thus a static domain-instability extension of ferroionic thermodynamics in which chemical capacitance governs whether the environment stabilizes a homogeneous state or allows a spatially modulated ferroelectric phase to emerge.

## II. Static problem and homogeneous thermodynamics

The film occupies spatial domain $0 < z < h$ as shown in Figure 1. The bottom electrode fixes $\phi(0) = 0$, while the upper surface is in equilibrium with a reservoir of species $X$. For an ideal gas, its chemical potential is

$$\mu_X(T, p_X) = \mu_X^\circ(T) + k_B T \ln(p_X/p^\circ) \tag{1}$$

The minimal scalar Landau free-energy density is

$$f_L(P,T) = \frac{1}{2}\alpha(T)P^2 + \frac{1}{4}\beta P^4 + \frac{1}{6}\gamma P^6, \qquad \alpha(T) = a(T - T_0) \tag{2}$$

with the homogeneous equation of state

$$E = f_L{}'(P) = \alpha P + \beta P^3 + \gamma P^5 \tag{3}$$

The second-order model has $\beta > 0$; the first-order model has $\beta < 0$ and $\gamma > 0$. These standard limits are used only to enable analytical derivations; the $BaTiO_3$ section below uses an eighth-order potential [40,41].

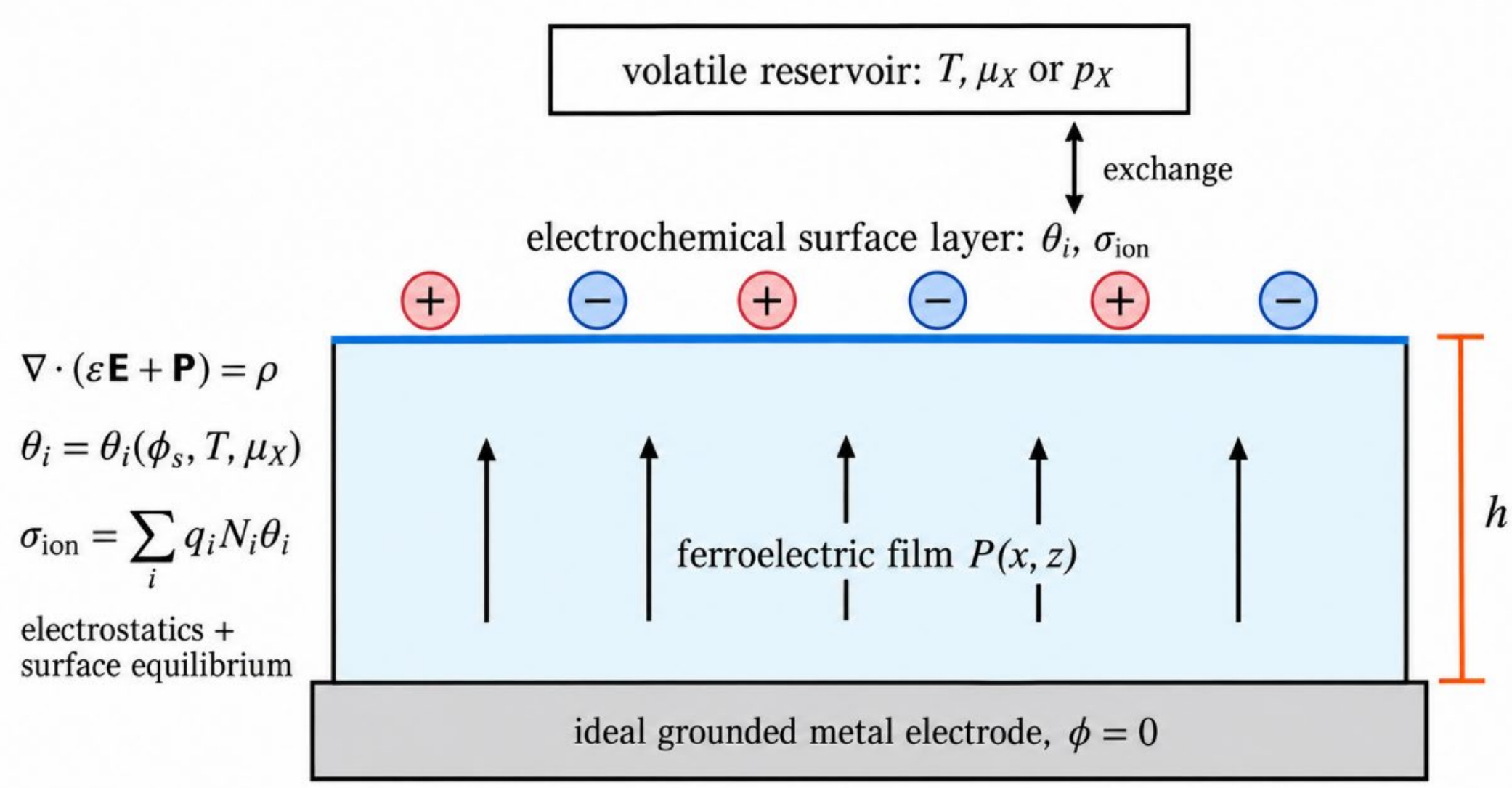

**Figure 1. Static geometry and variables.** A uniaxial ferroelectric film of thickness $h$ is grounded by an ideal bottom metal electrode and exposed at its upper surface to a volatile chemical reservoir. The surface populations $\theta_i$ generate the free charge $\sigma_{ion}$ and respond to the surface potential, temperature, and reservoir chemical potential. The analysis first treats a homogeneous polarization and then allows lateral modulation.

A surface species $i$ with charge $q_i$, site density $N_i$, standard formation free energy $\Delta G_i^0$, and stoichiometric coupling $\nu_i$ to the volatile reservoir has the Langmuir-type equilibrium

$$\frac{\theta_i}{1-\theta_i} = \exp\left[-\frac{\Delta G_i^0 - \nu_i \mu_X + q_i \phi_s}{k_B T}\right], \qquad \sigma = \sum_i q_i N_i \theta_i \tag{4}$$

This is the same thermodynamic structure used in electrochemical surface-compensation models [2,4]. More elaborate adsorption models change the functional form of $\sigma(\phi_s, T, \mu_X)$ but do not change the stability construction below.

For a homogeneous film, stationarity of the coupled ferroelectric–surface grand potential gives

$$E = f_L{}'(P), \qquad E = -\frac{P + \sigma(\phi_s, T, \mu_X)}{\varepsilon_f}, \qquad \phi_s = -hE \tag{5}$$

or, equivalently,

$$P + \varepsilon_f f_L{}'(P) + \sigma[-hf_L{}'(P), T, \mu_X] = 0 \tag{6}$$

Equation (6) defines the homogeneous ferroionic state and reduces to the Stephenson–Highland-type monodomain problem in the corresponding electrochemical limit [2,8].

The local response of the surface charge to potential is

$$C_{\text{chem}} = -\left(\frac{\partial\sigma}{\partial\phi_s}\right)_{T,\mu_X} \tag{7}$$

For mutually exclusive charge states on a common surface site, the general result is

$$C_{\text{chem}} = \frac{N_s}{k_B T}(\langle q^2\rangle - \langle q\rangle^2), \qquad \langle q^n\rangle = \sum_i q_i^n\,\theta_i \tag{8}$$

For independent Langmuir species occupying distinct site populations, Eq. (8) reduces to $C_{\text{chem}} = \sum_i q_i^2\,N_i\theta_i(1-\theta_i)/(k_B T)$.

Thus $C_{\text{chem}}$ is largest at partial occupation and decreases when a surface state is empty or saturated. This immediately distinguishes two quantities: $\sigma$ controls compensation of the existing mean polarization, while $C_{\text{chem}}$ controls the incremental ability of the chemical layer to follow a new electrostatic perturbation. The terminology also parallels the broader electrochemical concept of chemical capacitance as a differential storage response [42].

Linearizing the surface response as $\sigma \simeq \sigma_0 - C_{\text{chem}}\phi_s$ gives the reduced homogeneous equation

$$\left[\alpha + \frac{1}{\varepsilon_f + hC_{\text{chem}}}\right]P + \beta P^3 + \gamma P^5 = -\frac{\sigma_0}{\varepsilon_f + hC_{\text{chem}}} \tag{9}$$

The chemistry therefore has two distinct effects: it renormalizes the quadratic stiffness through $C_{\text{chem}}$ and produces a chemical bias field through the equilibrium charge $\sigma_0$. This separation is central to the domain problem.

## III. Instability of the homogeneous state

We perturb a homogeneous equilibrium $P_0$ by a columnar lateral mode $\delta P = p\cos(kx)$ and include the gradient energy $g(\nabla P)^2/2$. The local ferroelectric stiffness is

$$A_0 = f_L{}''(P_0) = \alpha + 3\beta P_0^2 + 5\gamma P_0^4 \tag{10}$$

Solving Laplace electrostatics in the ferroelectric and external region and linearizing the chemical boundary condition yields the inverse susceptibility of a mode $k$,

$$\Lambda(k) = A_0 + gk^2 + \left\{h\left[C_{\text{chem}} + k\left(\varepsilon_{\text{out}} + \varepsilon_f \coth kh\right)\right]\right\}^{-1} \tag{11}$$

The homogeneous state is stable when $\Lambda(k) > 0$ for every $k$. A domain-forming spinodal is reached when

$$\min_{k>0}\Lambda(k) = 0 \tag{12}$$

This expression makes the role of chemical capacitance explicit. A large $C_{\text{chem}}$ screens finite-$k$ fluctuations and stabilizes a homogeneous state; a small $C_{\text{chem}}$ leaves the polarization modulation electrostatically costly and favors the alternative of forming oppositely polarized domains.

For weak chemical response, $C_{\text{chem}} \to 0$, and $kh \gg 1$, Eq. (11) becomes

$$\Lambda(k) \simeq A_0 + gk^2 + \left[h\left(\varepsilon_f + \varepsilon_{\text{out}}\right)k\right]^{-1} \tag{13}$$

Minimization yields

$$k_c = \left[2gh\left(\varepsilon_f + \varepsilon_{\text{out}}\right)\right]^{-1/3}, \qquad L_c = 2\pi/k_c \propto h^{1/3} \tag{14}$$

The corresponding electrostatic-plus-gradient penalty scales as $h^{-2/3}$. The resulting critical-thickness relation follows directly by setting $\Lambda(k_c) = 0$ and is given explicitly in Supplementary Section S3. The $h^{1/3}$ scaling is the linear-instability result near onset and should not be confused with the $h^{1/2}$ Kittel scaling of fully developed sharp-wall domains.

For a stripe modulation $P = P_0 + A\cos(kx)$ about a finite homogeneous polarization $P_0$, the odd spatial moments vanish. The weakly nonlinear bulk free energy retains the same form with an effective quartic coefficient $\beta_{\text{eff}} \equiv \beta + 10\gamma P_0^2$:

$$\Delta f_s(A) = \frac{\Lambda}{4}A^2 + \frac{3\beta_{\text{eff}}}{32}A^4 + \frac{5\gamma}{96}A^6, \qquad \beta_{\text{eff}} \equiv \beta + 10\gamma P_0^2 \tag{15}$$

For $\beta_{\text{eff}} > 0$, the stripe amplitude grows continuously when $\Lambda$ becomes negative. For $\beta_{\text{eff}} < 0$, a finite-amplitude branch exists before the homogeneous state becomes linearly unstable. The stripe-state coexistence and branch-appearance thresholds are

$$\Lambda_{\text{coex}}^{s} = \frac{27\beta_{\text{eff}}^2}{160\gamma}, \qquad \Lambda_{\text{meta}}^{s} = \frac{9\beta_{\text{eff}}^2}{40\gamma} \quad (\beta_{\text{eff}} < 0) \tag{16}$$

Thus the pre-spinodal first-order stripe window exists only while $\beta_{\text{eff}} < 0$. For $\beta < 0$ and $\gamma > 0$, it closes at $P_0^2 = |\beta|/(10\gamma)$; above this polarization the stripe onset becomes continuous at $\Lambda = 0$.

For a normalized checkerboard perturbation about the same $P_0$, the quadratic term is identical and the same $\beta_{\text{eff}}$ enters the quartic term. Its first-order coexistence threshold is

$$\Lambda_{\text{coex}}^{c} = \frac{243\beta_{\text{eff}}^2}{1600\gamma} \quad (\beta_{\text{eff}} < 0) \tag{17}$$

Because $27/160 > 243/1600$, stripes preempt checkerboards whenever $\beta_{\text{eff}} < 0$. When $\beta_{\text{eff}} \geq 0$, both onsets become continuous at $\Lambda = 0$. Figure 2(b) shows the $P_0 = 0$ special case.

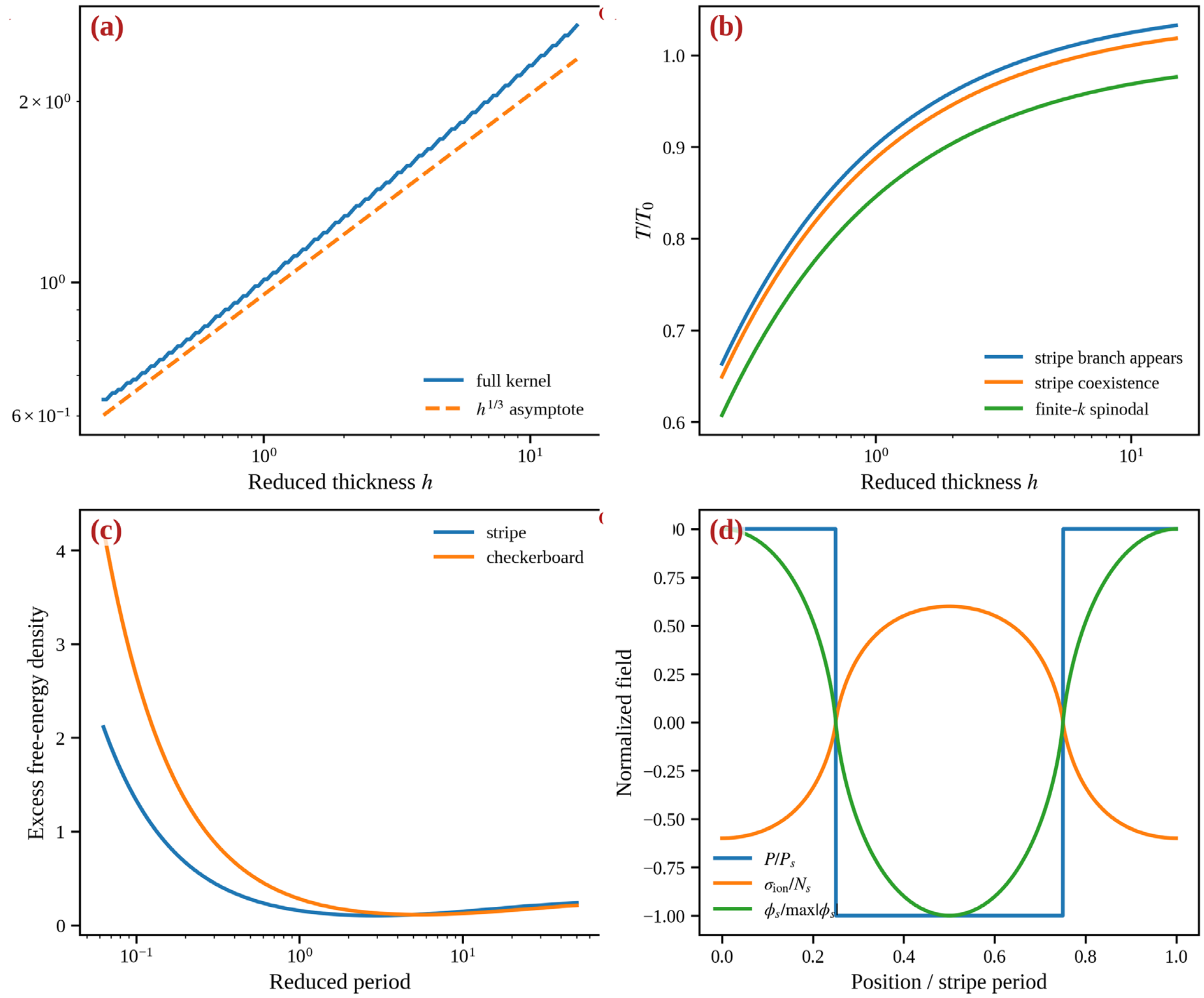

**Figure 2. Analytical structure of the domain instability.** (a) Selected finite-k onset period from the full electrostatic kernel compared with the $h^{1/3}$ asymptote. (b) First-order $P_0 = 0$ special case showing appearance of the metastable stripe branch, homogeneous-stripe coexistence, and the finite-k spinodal. (c) Full-Fourier sharp-wall comparison of stripe and checkerboard free energies. (d) Representative stripe state illustrating the spatial relation among polarization, ionic screening charge, and surface potential.

For developed stripes with 180° wall energy $\gamma_{\mathrm{DW}}$, the thick-film sharp-wall free energy can be written

$$F_s(L) = \frac{2\gamma_{\mathrm{DW}}}{L} + \frac{7\zeta(3)}{4\pi^3}\frac{P_s^2 L}{h(\varepsilon_f+\varepsilon_{\mathrm{out}})} \tag{18}$$

and therefore

$$L_s = \left[\frac{8\pi^3\gamma_{\mathrm{DW}} h(\varepsilon_f+\varepsilon_{\mathrm{out}})}{7\zeta(3)P_s^2}\right]^{1/2} \propto h^{1/2} \tag{19}$$

The two asymptotic laws therefore describe different regimes: $h^{1/3}$ at the onset of a smooth finite-$k$ instability and $h^{1/2}$ after sharp walls have developed.

Figure 3(a) and 3(b) show the checkerboard reference morphology and its induced surface-charge pattern used as the two-dimensional comparison state.

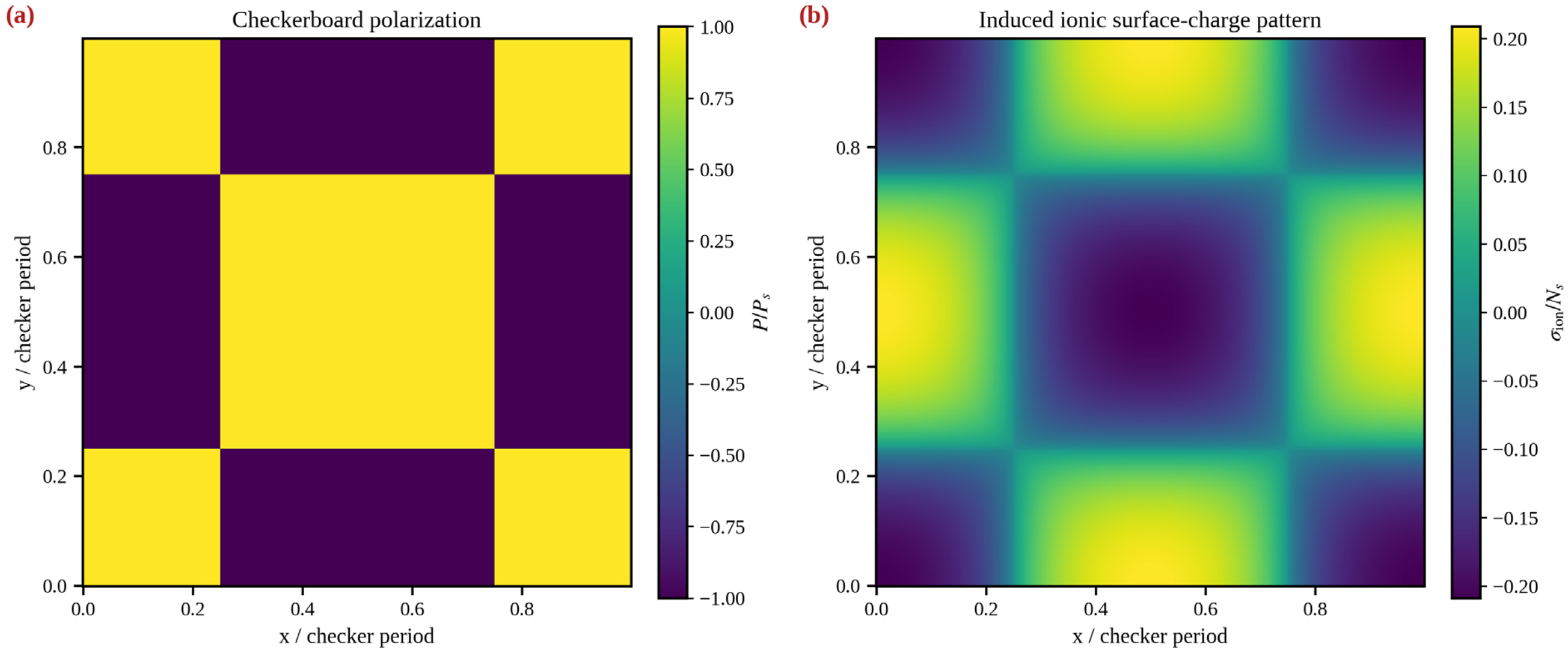

**Figure 3. Two-dimensional reference morphology.** (a) Checkerboard polarization used as a competing two-dimensional modulation. (b) Induced ionic surface-charge pattern. Electrostatic nonlocality smooths the chemical response near wall intersections; the checkerboard remains higher in free energy than stripes for the scalar isotropic model.

## IV. Reduced numerical phase structure

The analytical results establish the boundaries and scalings but do not by themselves show how temperature, chemical potential, and thickness reorganize the complete set of homogeneous and modulated states. We therefore solved the coupled homogeneous electrochemical equation and evaluated the finite-$k$ and weakly nonlinear domain criteria on dense two-dimensional grids. The reduced parameters and convergence details are collected in Supplementary Section S2. The calculations deliberately use a dimensionless symmetric three-state surface chemistry so that the topology can be examined independently of a specific material.

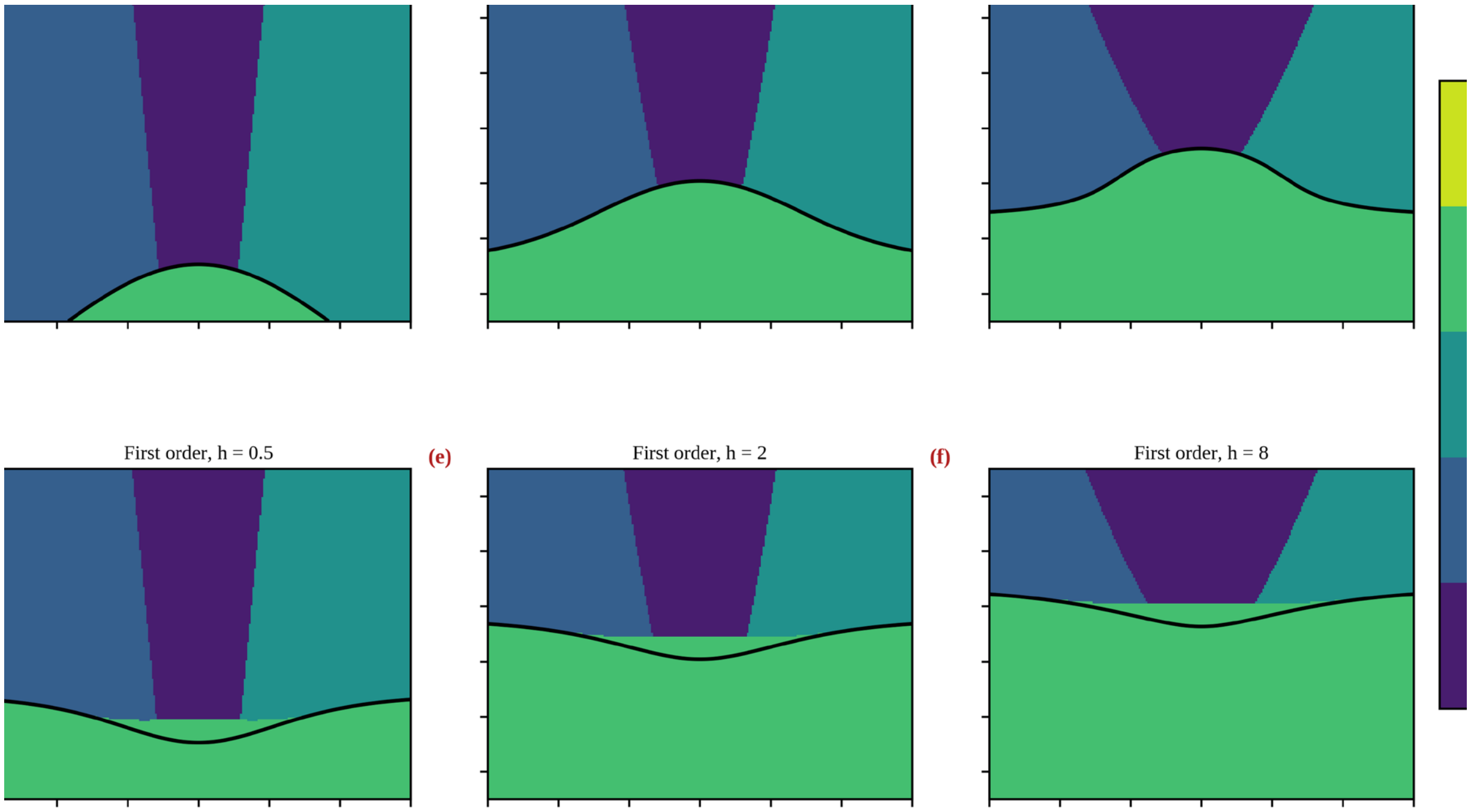

**Figure 4. Dense reduced temperature-chemical-pressure phase atlas.** (a)-(c) Second-order ferroelectric for reduced thickness h = 0.5, 2, and 8. (d)-(f) First-order ferroelectric at the same thicknesses. Colored fields classify the lowest state within the weakly nonlinear variational comparison, whereas the thick black curve is the local finite-k spinodal $\Lambda_{min} = 0$. The green stripe-favored region and black spinodal coincide for the continuous second-order onset to numerical resolution, but separate near chemical neutrality in the first-order case because a finite-amplitude stripe state can have lower free energy while the homogeneous state remains metastable.

The analysis of these results yields three robust trends. First, the central chemical-pressure region is where domain formation is most competitive because the environment does not strongly bias one polarization orientation. Second, increasing thickness systematically expands the stripe-favored region because the wall cost per unit volume decreases relative to the depolarization-energy benefit. Third, the order of the bulk ferroelectric transition changes the nature of the boundary: a second-order system enters the stripe state continuously at the finite-$k$ instability, whereas a first-order system can enter it by finite-amplitude nucleation while the homogeneous state remains locally stable. The colored classification and the black curve encode different mathematical criteria: the color is a finite-amplitude thermodynamic comparison among the retained ansatz states, while the black curve is a local linear spinodal. Their separation in the center of the first-order diagrams is therefore physical within the model rather than a plotting inconsistency; toward the chemically biased flanks the finite-amplitude advantage weakens and the two criteria approach one another. In the finite-$P_0$ expansion this crossover is controlled by the effective quartic coefficient $\beta + 10\gamma P_0^2$: the discontinuous stripe window closes when this coefficient changes sign, i.e. at $P_0^2 = |\beta|/(10\gamma)$ for $\beta < 0$.

The distinction between equilibrium screening charge and chemical capacitance is visible most directly when the local stability and the selected wavelength are plotted separately. Figure 5(a)–(f) separates the local stability margin, selected wavelength, differential chemical response, and schematic state interpretation for the reduced model.

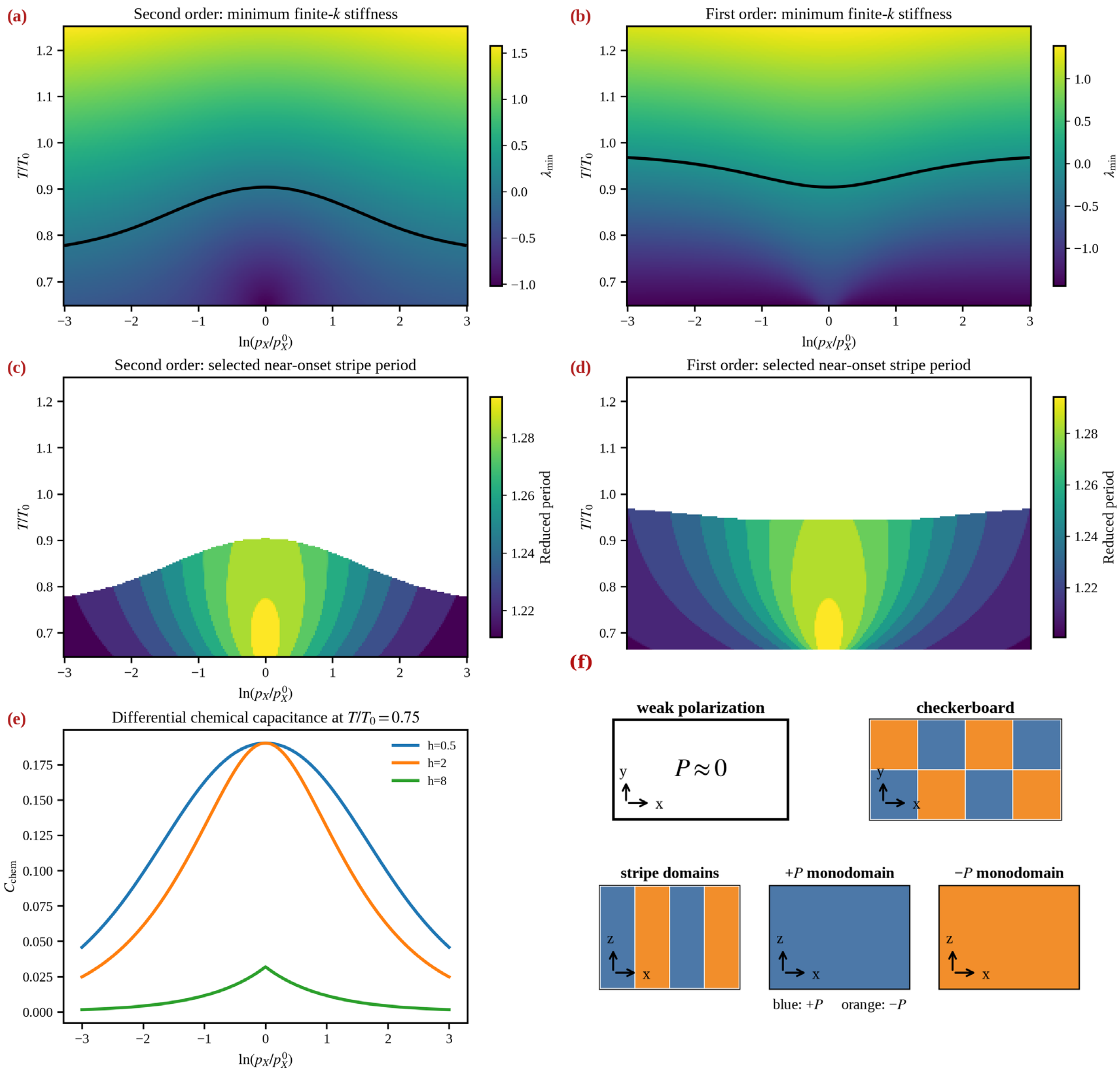

**Figure 5. Stability measures and state interpretation for the reduced model.** (a,b) Minimum finite-k stiffness for second- and first-order ferroelectrics at $h$ = 2; the thick black contour marks $\Lambda_{min}$ = 0. (c),(d) Selected near-onset stripe period. (e) Differential chemical capacitance versus chemical pressure for three thicknesses. (f) Schematic states: weak polarization and checkerboard configurations are shown in the x-y plane on the top row, whereas stripe, +P monodomain, and −P monodomain configurations are shown in x-z cross-section on the bottom row.

The domain wavelength varies much less strongly with chemical pressure than the phase boundary itself. In the developed-domain calculation, chemical bias is accommodated primarily by changing the fraction of the two polarization variants, i.e. by moving walls, before it strongly changes the preferred period. This suggests that experimentally chemical potential should affect domain fraction and net polarization more strongly than domain density over a broad part of the polydomain regime.

**V. Dimensional realization for $BaTiO_3$**

To place the theory on physical temperature and thickness scales, we use the eighth-order $BaTiO_3$ potential of Li, Cross, and Chen [41], restricted to the tetragonal out-of-plane order parameter,

$$f_{\mathrm{BTO}} = \alpha_1(T)P^2 + \alpha_{11}P^4 + \alpha_{111}P^6 + \alpha_{1111}P^8 \quad (20)$$

The coefficients, dielectric background, and two gradient-energy calibrations are given in Supplementary Table S2. This potential yields a cubic–tetragonal coexistence temperature near 398.15 K, a polarization jump of approximately 0.181 C $m^{-2}$, and $P_s \approx 0.259$ C $m^{-2}$ at 300 K. We use a literature background dielectric constant and gradient coefficient [43] and also show a sensitivity calibration tied to the lower first-principles range of $BaTiO_3$ 180° wall energies [44–46]. Independent first-principles effective-Hamiltonian calculations reproduce the first-order $BaTiO_3$ transition sequence and provide a complementary microscopic benchmark [47].

The oxygen surface chemistry is intentionally treated as an effective Stephenson–Highland-type boundary model rather than as a universal $BaTiO_3$ material constant. The numerical benchmark uses oppositely charged oxygen-related surface states with a common effective formation energy of 0.20 eV and a site area of 1 $nm^2$. These values are taken as model parameters, whereas actual boundaries will depend on surface termination, defects, water activity, and preparation history. The purpose of the $BaTiO_3$ section is therefore to establish dimensional scales and sensitivity, not to claim a unique termination-independent T–$pO_2$ phase diagram. First-principles surface calculations further show that $H_2O$ adsorption, hydroxylation, strain, and surface termination can alter the electrochemical coupling to $BaTiO_3$ polarization [14].

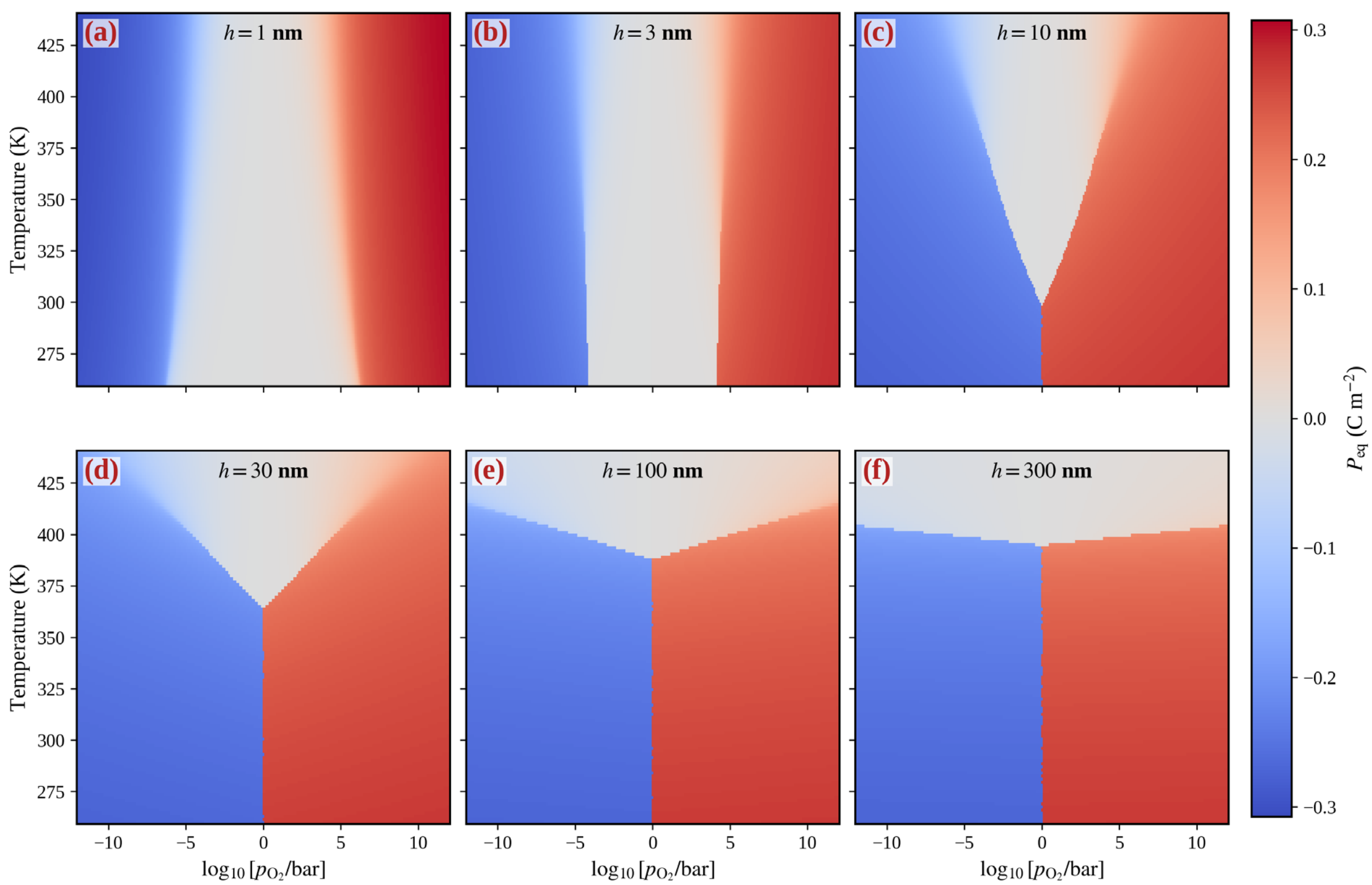

**Figure 6. $BaTiO_3$ homogeneous equilibrium polarization under the effective oxygen-redox boundary condition.** (a)-(f) Film thicknesses 1, 3, 10, 30, 100, and 300 nm, plotted on a common polarization scale in real temperature and oxygen pressure. Increasing thickness progressively restores bulk-like polarization while retaining chemically controlled polarization selection.

The six thicknesses span an ultrathin regime in which depolarization and chemical compensation dominate, an intermediate regime in which polarization and surface chemistry compete strongly, and a thick-film regime approaching bulk-like polarization magnitude. The sharp features in the maps are equilibrium branch changes of the homogeneous model rather than domain-instability boundaries. Figure 6(a)–(f) shows the corresponding equilibrium polarization maps across the six film thicknesses.

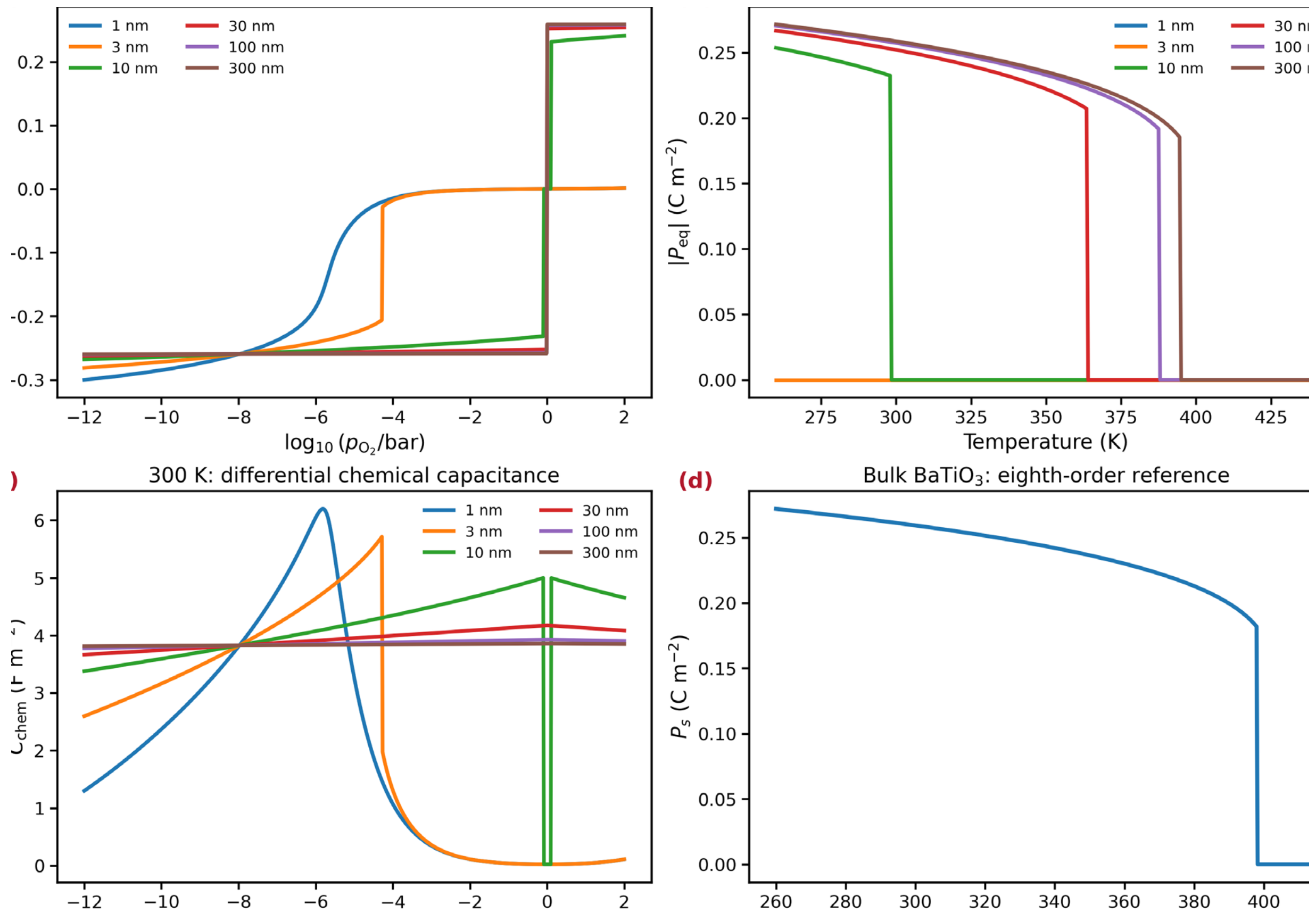

**Figure 7. $BaTiO_3$ equilibrium summaries.** (a) Equilibrium polarization versus oxygen pressure at 300 K. (b) Absolute equilibrium polarization versus temperature at 1 bar O2. (c) Differential chemical capacitance versus oxygen pressure at 300 K, recomputed by dense polarization bracketing, Brent refinement of every stationary root, and explicit global grand-potential selection. The 100 and 300 nm curves are smooth; the oscillations in the earlier plot were numerical branch-selection artifacts. (d) Bulk spontaneous polarization from the eighth-order $BaTiO_3$ thermodynamic potential, included as the dimensional reference.

Figure 7(c) emphasizes the main physical distinction of the paper. A thick film can have a chemically selected, nearly bulk-magnitude polarization while the differential surface response is relatively modest. Conversely, in partially occupied surface regimes the chemical capacitance can become large and strongly stabilize the homogeneous state against incremental modulations. Figure 7(d) provides the bulk reference needed to distinguish surface-chemical size effects from the intrinsic first-order $BaTiO_3$ transition.

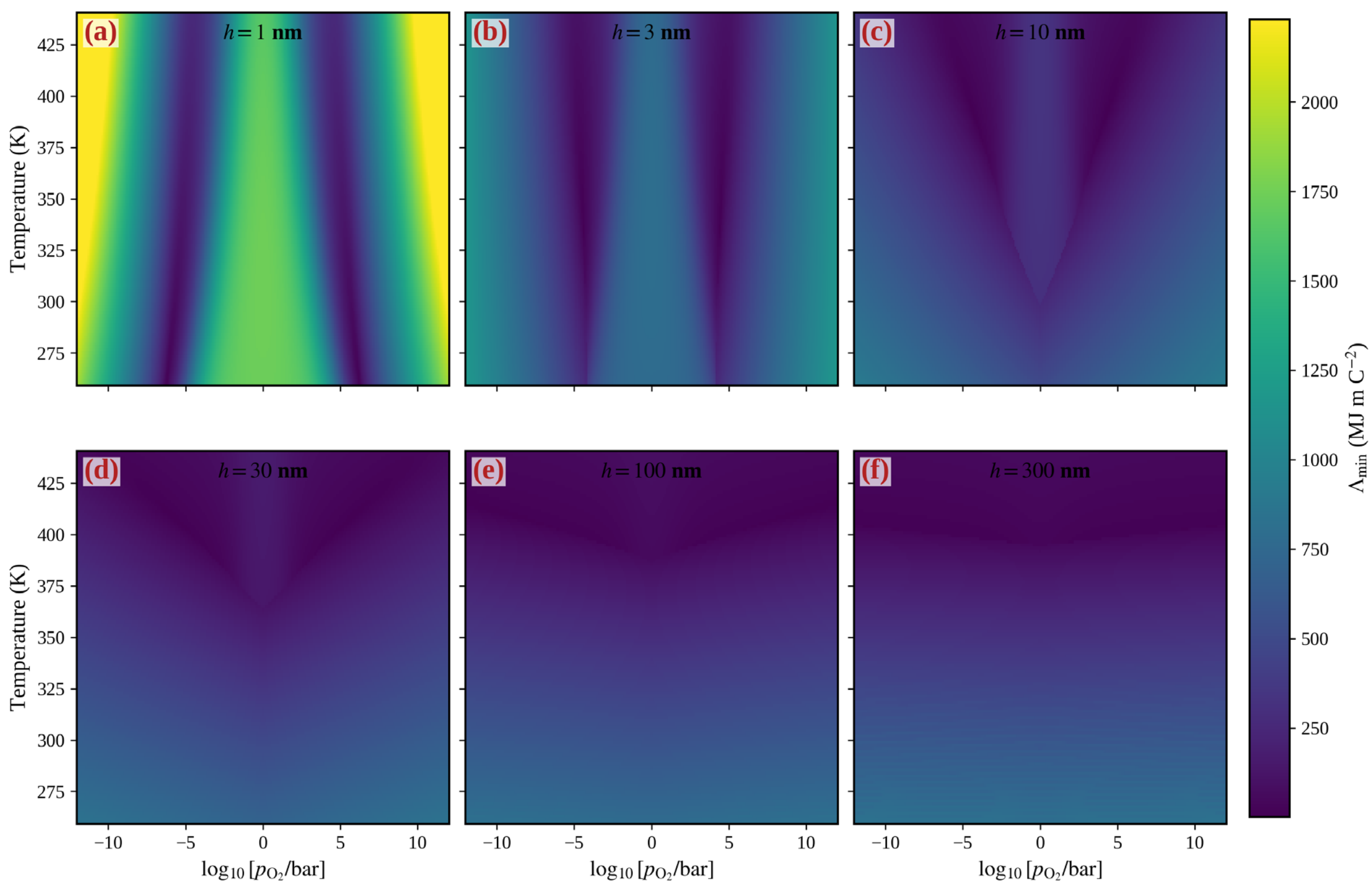

**Figure 8. $BaTiO_3$ annealed-ion finite-k stability margin**. (a)-(f) correspond to 1, 3, 10, 30, 100, and 300 nm films under the effective oxygen-redox boundary condition. The plotted quantity is $\Lambda_{min}$. It remains positive throughout the displayed window within the columnar 180° mode family; therefore no zero-instability contour occurs within that restricted subspace. A common display scale is used and clipped only at the 98.5th percentile to keep all six thicknesses legible.

For the chosen effective oxygen chemistry, the homogeneous equilibrium branches are locally stable against the columnar 180° finite-$k$ mode family across the plotted temperature–pressure range. This should not be interpreted as a universal prediction that $BaTiO_3$ free surfaces remain monodomain. Rather, it shows how a large chemical response can suppress this restricted domain instability. Reducing available surface charge, reducing $C_{\mathrm{chem}}$, or changing termination chemistry shifts the system toward the ion-poor limit. A dynamically frozen ionic layer is distinct: it retains the mean charge established during annealing while $\delta\sigma = 0$ and $C_{\mathrm{chem}}^{\mathrm{dyn}} = 0$, so its background $P_0$ and stability kernel need not coincide with the ion-poor case. The absence of a zero contour is a result within the columnar 180° subspace, not an omission; the display clipping in Figure 8(a)–(f) affects only the upper color range and never the sign of $\Lambda_{\mathrm{min}}$. Three-dimensional ferroionic calculations have likewise found that mono- and polydomain stability depends strongly on surface-ion formation energy, thickness, voltage, and temperature [48].

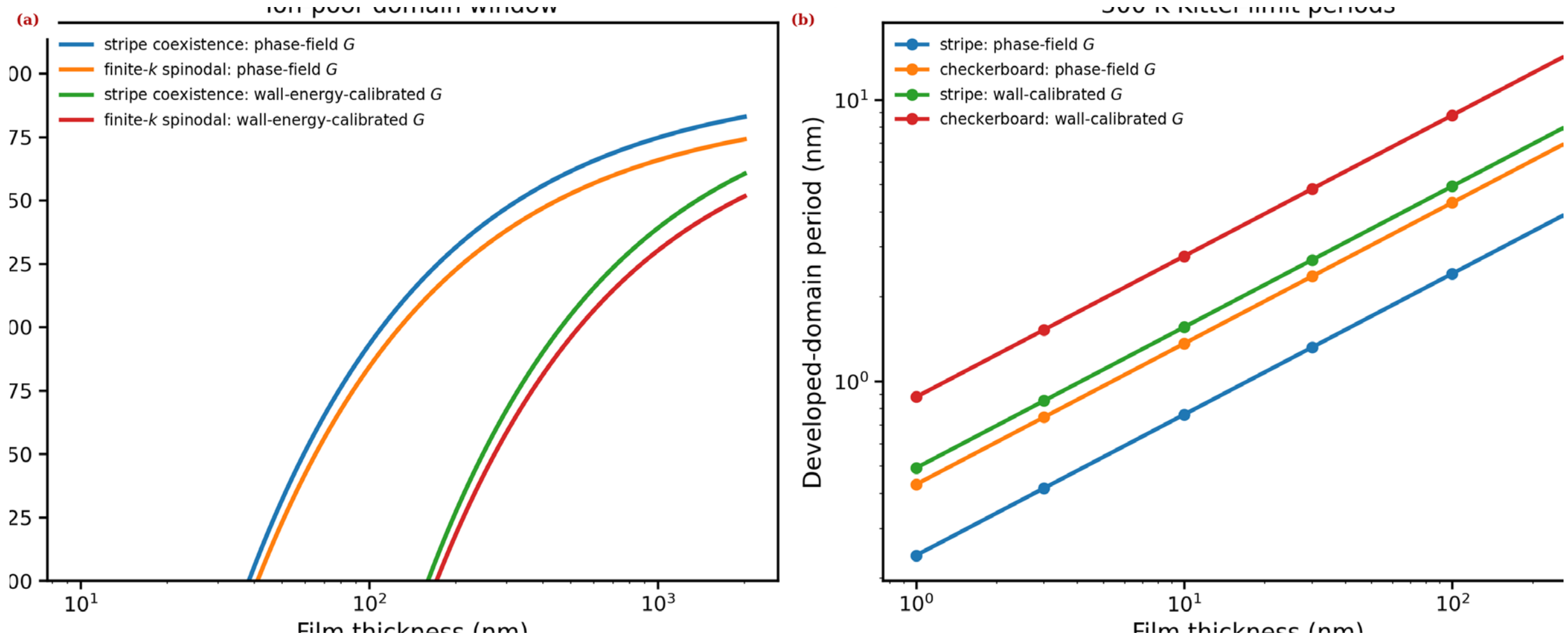

**Figure 9. $BaTiO_3$ domain scales when chemical screening is ineffective.** (a) Stripe coexistence and finite-k spinodal in the ion-poor/open-circuit limit for two gradient-energy calibrations. (b) Developed-domain Kittel-limit stripe and checkerboard periods at 300 K for 1-300 nm films. The gradient calibrations expose the quantitative sensitivity to wall energy, while stripes remain favored over checkerboards.

At 300 K, the literature phase-field gradient coefficient gives a stripe-coexistence thickness of approximately 111 nm and a finite-$k$ spinodal near 128 nm. The wall-energy-calibrated gradient coefficient shifts these values to approximately 463 and 534 nm, respectively. The large spread is physically important: the instability is highly sensitive to the gradient and wall-energy scale, and quantitative predictions require a parameterization appropriate to the actual film, strain state, and wall structure. Nevertheless, the qualitative result is robust: annealed chemical screening and ion-poor/open-circuit conditions can place the same nominal material on opposite sides of the monodomain–polydomain boundary. Figure 9(a) and 9(b) summarize the ion-poor domain window and the corresponding Kittel-limit periods for the two gradient calibrations.

## VI. Chemical control of domain fractions in the polydomain state

The instability analysis determines whether a homogeneous ferroionic state gives way to a spatially modulated state, but it does not determine how an established polydomain state partitions between $c^+$ and $c^-$ variants. That second problem is controlled primarily by the zero-wave-vector chemical bias. Let $\eta=f^+$ and $1-\eta=f^-$ be the fractions of two stripe variants with polarizations $P_p$ and $P_m$. The mean polarization is

$$\bar{P} = \eta P_p + (1-\eta) P_m. \tag{21}$$

For a sharp stripe pattern of period L, the nonzero Fourier amplitudes are

$$|P_n|^2 = \frac{(P_p - P_m)^2}{\pi^2 n^2} \sin^2(\pi n \eta). \tag{22}$$

The free-energy density separates naturally into a zero-mode term $F_0$ that contains the bulk and uniform electrochemical contributions, the domain-wall cost, and the finite-wave-vector electrostatic terms:

$$F(L,\eta) = F_0(\eta) + \frac{2\gamma_w}{L} + \sum_{n=1}^{\infty} |P_n|^2 K_n. \tag{23}$$

The finite-wave-vector screening kernel is

$$K_n = \left[h\left(C_n + k_n\left(\epsilon_{out} + \epsilon_f \coth(k_n h)\right)\right)\right]^{-1}, \qquad k_n = \frac{2\pi n}{L}. \tag{24}$$

Thus chemical bias enters the domain-fraction problem through the zero mode, whereas the period and energetic cost of the spatial modulation are controlled by the finite-k harmonics. This distinction complements the stability analysis above: the latter determines whether a modulated state is favored, while the present calculation determines how that state is partitioned between the two polarization variants.

For symmetric 180° domains, $P_p=+P_s$ and $P_m=-P_s$, define the imbalance $m=2\eta-1=f^{+}-f^{-}$. In the developed-domain, weak-incremental-screening limit, $C_n\to 0$ and $knh\gg 1$, minimization over the stripe period gives

$$m = 2\eta - 1, \qquad L_{opt}(m) = \left[\frac{\gamma_W \pi^3 h \epsilon_{sum}}{P_s^2 S_3(m)}\right]^{1/2}. \tag{25}$$

where the fraction-dependent Fourier sum is

$$S_3(m) = \sum_{n=1}^{\infty} \frac{\sin^2[\pi n(1+m)/2]}{n^3}. \tag{26}$$

Near equal fractions, the optimized period has the expansion

$$\frac{L_{opt}(m)}{L_0} = 1 + \frac{\pi^2 \ln 2}{7\zeta(3)} m^2 + O(m^4) \simeq 1 + 0.813 m^2. \tag{27}$$

Equation (27) gives a useful qualitative result: the domain fraction changes linearly with a weak chemical bias, whereas the optimized period changes only quadratically. The first response of a stable stripe state to chemistry is therefore wall translation and redistribution of $c^+$ and $c^-$ widths, rather than a large change in domain density.

If the mean surface response is linearized around a chemically prepared state, with inherited surface charge $\sigma_0$ and zero-mode differential capacitance $C_0$, the mean contribution and thick-film limiting relation are

$$F_{mean}(m) = \frac{(P_s m + \sigma_0)^2}{2(\epsilon_f + hC_0)}, \qquad m \simeq -\frac{\sigma_0}{P_s}, \qquad f_p = \frac{1+m}{2}, \quad f_m = \frac{1-m}{2}. \tag{28}$$

In the thick-film limit the wall and finite-k correction becomes small and Eq. (28) reduces to a ferroionic lever rule: the domain imbalance compensates the inherited mean surface charge. As |m| approaches unity the minority-domain fraction vanishes and Eq. (25) drives the equilibrium period to infinity, producing a chemically selected monodomain.

This continuous fraction tuning should not be confused with the unconstrained grand-canonical equilibrium of a macroscopically coarsened, fully annealed surface. If each macroscopic domain can independently exchange ions with the reservoir and the period can grow without constraint, the lower-grand-potential polarization orientation is selected away from coexistence. Continuous pressure-dependent fractions are therefore most directly relevant to finite-period states and to situations with frozen or conserved surface charge, slow lateral ionic redistribution, wall pinning, or finite sample size. This is the same physical distinction emphasized above between annealed and frozen incremental screening.

We illustrate the fraction thermodynamics for $BaTiO_3$ at 300 K using the same eighth-order bulk potential, effective symmetric oxygen-redox boundary model, and wall-energy calibration as in Section V. Because the sharp-wall construction is not appropriate for the 1 and 3 nm limits, Figure 10 shows 10, 30, 100, and 300 nm films. The oxygen pressure sets the inherited mean charge σ0 of the prepared surface, while the finite-k ionic response is taken as frozen during domain redistribution. This construction isolates the thermodynamic tendency of mobile walls to change the $c^+/c^-$ fraction under a chemically inherited bias.

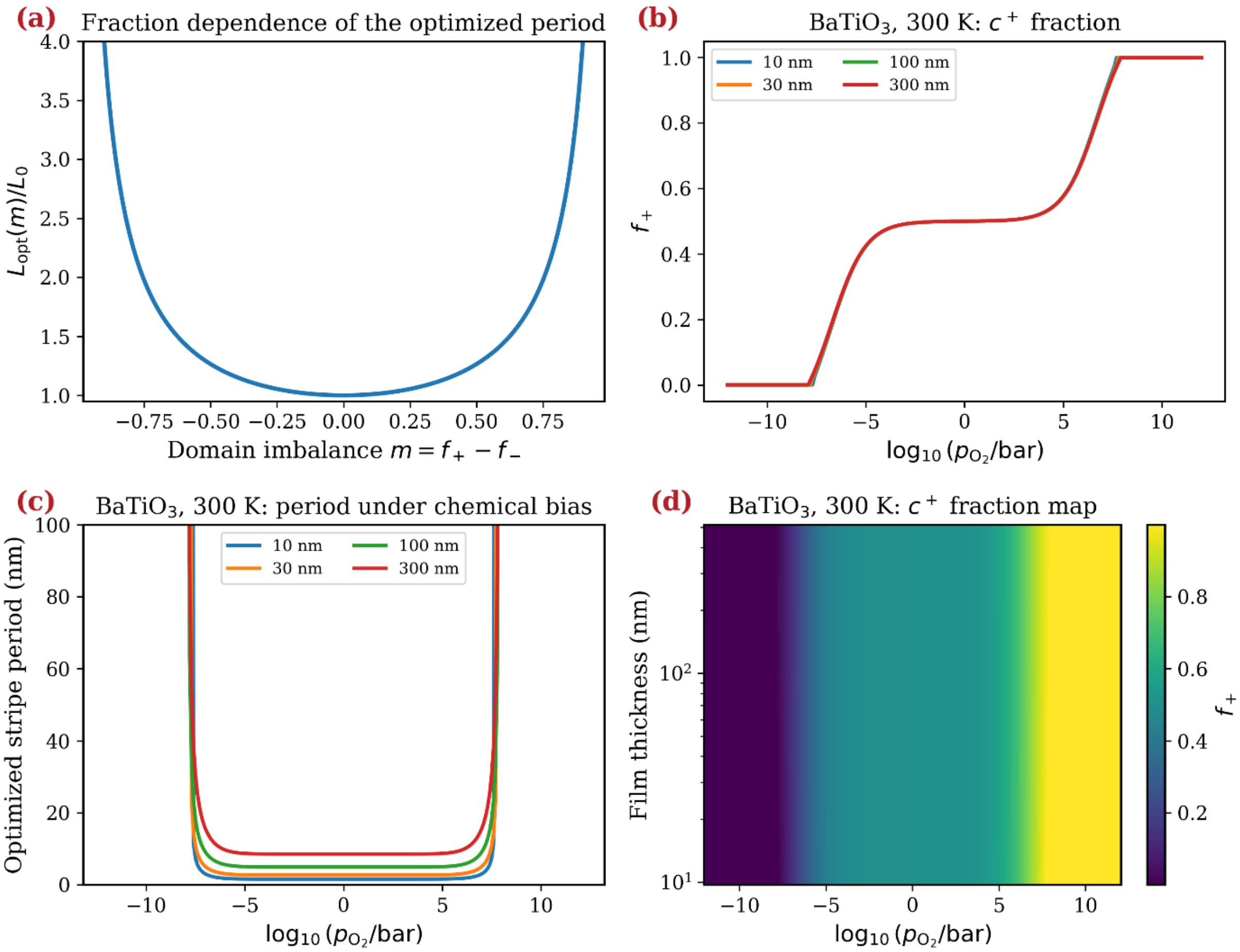

**Figure 10. Chemical control of domain fractions in developed 180° stripe states.** (a) Universal dependence of the optimized stripe period on domain imbalance $m=f^+-f^-$ in the developed-domain limit; the period changes only quadratically near equal fractions. (b) $BaTiO_3$ $c^+$ fraction versus the effective oxygen-pressure coordinate at 300 K for 10, 30, 100, and 300 nm films in the frozen-incremental-screening construction. (c) Corresponding optimized stripe period. (d) Thickness–pressure map of the $c^+$ fraction. The fraction is much more sensitive to chemical bias than the period over the central polydomain regime, whereas the period diverges as the minority domain disappears.

For the symmetric effective surface chemistry, $f^+=f^-=1/2$ at chemical neutrality. Increasing the oxygen chemical potential favors negative surface charge and therefore increases the $c^+$ fraction; the opposite pressure direction favors $c^-$. The fraction curves are only weakly thickness dependent because the leading balance is between mean polarization and inherited chemical charge, while the absolute period retains the strong thickness dependence of the domain-wall/electrostatic competition. Figure 10 therefore completes the static thermodynamic picture: the instability analysis determines which spatial phase is available, and the fraction analysis determines the internal composition of that phase once domains exist.

## VII. Discussion and outlook

The present theory now separates three related static questions: homogeneous electrochemical thermodynamics, instability toward a modulated state, and the fraction thermodynamics of an established 180° stripe state. All three remain within a scalar out-of-plane framework. The instability analysis assumes columnar 180° modulations and therefore excludes polarization rotation, ferroelastic 90° variants, closure domains, vortices, elastic compatibility, and flexoelectricity. The weakly nonlinear phase maps compare a restricted set of stripe and checkerboard ansatz states rather than an unconstrained two- or three-dimensional phase-field minimization. The domain-fraction treatment adds a second restriction: Eqs. (21)–(26) describe developed sharp-wall stripes, and the continuous $BaTiO_3$ fraction curves in Figure 10 are evaluated in the frozen-incremental-screening limit. A fully annealed grand-canonical surface with unconstrained coarsening instead selects the lower-grand-potential monodomain away from coexistence. Thus the fraction results apply most directly to finite-period, pinned, conserved-charge, or dynamically frozen polydomain states.

Other limitations remain those of the chemical and electrostatic boundary models. The surface chemistry is represented by local Langmuir-type equilibria, while real surfaces may contain multiple site types, lateral interactions, defect equilibria, water-mediated reactions, electronic carriers, and reconstruction. Slow ions can interpolate between the annealed and frozen limits considered here, and lateral conservation can make the effective screening response wave-vector dependent. The quantitative $BaTiO_3$ surface parameters are illustrative because a unique termination-independent oxygen adsorption free energy and active site density do not exist. The electrostatic model also assumes an ideal bottom electrode and zero separation between the polarization bound-charge plane and the ionic-charge plane; finite electrode screening, a dead layer, or finite adsorbate standoff can be important in the 1–3 nm regime. Mechanical boundary conditions can shift transition temperatures and phase sequences, and the capillarity nucleation construction in the Supplementary Material should not be interpreted as the microscopic switching pathway in defect-containing films. These restrictions do not alter the principal analytical separation established here: the differential chemical response controls the onset of spatial instability, while the mean chemical bias controls the fraction of the polarization variants within a stable polydomain state.

**VIII. Predictions and connection to prior work**

The combined instability and fraction theory suggests several falsifiable tests. First, changing chemical activity should shift not only the average polarization but also the boundary between homogeneous and polydomain states; the strongest monodomain stabilization should correlate with a large differential chemical capacitance rather than simply with the largest net surface charge. Second, once a stripe state exists, a weak chemical bias should change the $c^+/c^-$ fraction to first order while changing the period only to second order, as expressed by Eq. (25). Domain walls should therefore translate and the net polarization should change before the domain density changes appreciably; at stronger bias the minority fraction should vanish and the period should diverge. Third, near the continuous finite-k onset the modulation wavelength should follow the $h^{1/3}$-type scaling, whereas developed sharp-wall patterns should cross toward Kittel-like behavior. Fourth, first-order ferroelectrics should display a pre-spinodal finite-amplitude stripe region only while the effective quartic coefficient is negative, so the separation between equilibrium domain appearance and the homogeneous spinodal closes as the biased background polarization increases. Fifth, chemically active and chemically frozen measurements on the same film should produce different stability boundaries and different fraction responses because the former permits incremental ionic screening whereas the latter retains the inherited mean charge with vanishing fast chemical response.

These predictions can be tested by combining controlled oxygen pressure or humidity with PFM, KPFM/surface-potential imaging, and temperature-dependent domain measurements. The most direct test of the new fraction result is a pressure or chemical-potential sweep while measuring both the areal $c^+/c^-$ fraction and the stripe period: the theory predicts a substantially stronger initial change in fraction than in period. Synchrotron scattering provides a complementary route for determining the average polarization and transition boundary, while surface-potential measurements constrain the screening state. Together these measurements can separate the equilibrium charge that biases the domain fractions from the differential chemical response that enters the domain-instability kernel.

## IX. Summary and outlook

The static thermodynamics of a chemically compensated ferroelectric involves two distinct chemical quantities and two distinct domain questions. The differential chemical capacitance determines how effectively the surface can follow a developing finite-wave-vector fluctuation and therefore controls the stability of the homogeneous ferroionic state. The equilibrium or inherited mean surface charge instead biases the average polarization and, once a polydomain state exists, controls the relative fractions of the oppositely polarized variants.

The resulting hierarchy therefore reveals the evolution of the ferroelectric domain structures in these systems. The instability analysis determines whether the system is homogeneous or modulated and which morphology is favored; the fraction analysis determines the internal composition of the modulated phase. Near equal $c^+/c^-$ fractions, chemical bias changes the fraction linearly while the optimized period changes only quadratically, so wall motion precedes substantial reconstruction of the domain density. Film thickness continues to control the wall/electrostatic balance and the absolute domain period, while the order of the bulk transition and the background polarization determine whether the domain phase appears continuously or through a finite-amplitude pre-spinodal branch. In the scalar isotropic model, stripes remain favored over checkerboards in the limits analyzed here.

The $BaTiO_3$ realization illustrates both layers of this thermodynamics. Under an annealed, strongly responsive effective oxygen-redox surface, the homogeneous state can remain locally stable against the columnar 180° modes considered here, whereas ion-poor conditions produce a thickness-dependent domain instability. Once a finite-period stripe state is present under a frozen or constrained chemical bias, the same surface chemistry shifts the $c^+/c^-$ fractions much more strongly than the period before driving the minority domain to extinction. Surface chemistry can therefore control not only whether domains form, but also the composition of the domain state after it forms. Extending the framework to laterally conserved ions, explicit reaction kinetics, vector polarization, elastic coupling, and unrestricted multidimensional microstructures provides the natural route from the present static theory to fully coupled ferroionic domain dynamics.

**Author and AI contributions:** Sergei V. Kalinin conceived the overarching scientific problem of coupling surface electrochemistry to ferroelectric phase stability, specified the electrochemical boundary conditions and the comparison of first- and second-order ferroelectrics, directed the analysis of domain formation as functions of temperature, chemical potential, and thickness, selected the $BaTiO_3$ dimensional realization, and introduced the questions of domain morphology, nucleation, and pressure-dependent $c^+/c^-$ domain fractions. OpenAI Astra assisted with literature organization, proposed and developed the use of differential chemical capacitance as the incremental screening variable, carried out algebraic derivations and asymptotic analyses of the finite-wave-vector instability and domain-fraction thermodynamics, implemented the numerical calculations and figures in Python, and assisted with manuscript drafting and revision under the author's direction. Fable 5.1 was used as an independent checker of selected analytical derivations and manuscript consistency. The author reviewed, corrected, and approved the physical assumptions, calculations, interpretation, and final text.

**Acknowledgements:** This work (SVK) was supported by the DOE BES project DE-SC0026253, Deciphering Electrochemical Transformation Pathways on the Nanometer Scale: Advancing Fundamental Discovery for Material Innovation

**Supplementary Material**

## S1. Detailed historical development of polarization–surface-chemistry coupling

### S1.1. 1940s–1950s: Landau–Devonshire theory, domains, catalysis, etching, and surface space charge

The theoretical language used in the main text originates in the phenomenological expansion of the ferroelectric free energy developed for $BaTiO_3$ by Devonshire [40]. In parallel, the energetic logic of domain formation inherited the wall-energy-versus-long-range-field competition formalized by Kittel for ferromagnets [35] and adapted to ferroelectrics by Mitsui and Furuichi [36]. These developments established two ingredients that remain central here: a local order-parameter free energy and a nonlocal electrostatic incentive to form domains.

The earliest explicit connection between ferroelectric order and surface chemical reactivity is generally traced to Parravano in 1952 [16]. The experiment examined catalytic oxidation near ferroelectric transitions of niobates and reported anomalous reaction behavior associated with the transition. The microscopic origin was not established, but the proposition that a bulk ferroelectric state can modify a heterogeneous reaction was already present.

Hooton and Merz provided a more direct polarization-resolved observation in 1955 [17]. $BaTiO_3$ single crystals etched in hydrochloric acid exhibited a strongly different etch rate at the positive and negative ends of the spontaneous polarization. Although the original purpose was domain identification, the result directly demonstrated that polarization sign changes a surface reaction rate.

At essentially the same time, Känzig [18] and Chynoweth [19] made surface charge itself a central physical object. Känzig introduced the concept of a space-charge layer near a ferroelectric surface, while Chynoweth inferred surface space-charge layers in $BaTiO_3$ from pyroelectric, photovoltaic, and hysteresis observations and argued that these fields could influence polarization orientation, apparent transition temperature, and domain nucleation. These papers did not chemically identify the compensating species, but they established that a ferroelectric surface has its own charge state and that this state feeds back on bulk polarization.

Merz's observations of domain nucleation and wall motion [51] and Landauer's electrostatic analysis of a reversed nucleus [49] established the classical switching problem. Landauer's calculation exposed the nucleation paradox: plausible wall and electrostatic energies predict homogeneous nucleation barriers much larger than experimentally inferred values. The present capillarity treatment in Section S4 should therefore be understood as a thermodynamic limiting construction rather than a microscopic nucleation mechanism.

### S1.2. 1960s–1990s: atmosphere-dependent switching and polarization-controlled adsorption/catalysis

Toyoda and Itakura reported in 1962 that polarization reversal in $BaTiO_3$ with gold electrodes slowed on evacuation and accelerated after adsorption of polar molecules, with the atmosphere at the anode being particularly important [20]. In modern terminology, this is an early observation of environment-dependent polarization kinetics. Kay and Dunn's thickness-dependent nucleation measurements [52] and Stadler's demonstration that ferroelectric polarization charging can alter supported-metal properties [53] further emphasized the role of boundaries and adjacent materials.

Cabrera, Sales, Maple, Suhl, and coworkers brought direct surface-science measurements to the problem in 1979 [21]. Auger spectroscopy showed selective adsorption of NO on $KNbO_3$ domains of a particular polarization orientation, and temperature-programmed desorption resolved multiple adsorption states. Importantly, the same work found no simple catalytic-rate anomaly for every reaction examined, demonstrating that polarization-dependent adsorption does not imply a universal catalytic response.

Kretschmer and Binder developed a phenomenological Landau treatment of free-surface effects on ferroelectric phase transitions [54], an important precursor to later finite-size LGD theories. In the 1980s and early 1990s, Inoue and collaborators developed a sustained program using poled $LiNbO_3$ as an active support for metals and semiconducting oxides [22–24]. Polarization-dependent CO oxidation, adsorption-induced conductivity changes, and desorption energetics were interpreted through electrostatic fields and band bending. Their 1993 discussion of device-type catalysts made explicit the concept of controllable catalytic activity [24].

By 2000, four elements of the present framework were therefore already known separately: polarization-dependent chemistry, environmental control of ferroelectric behavior, surface/space charge as an independent state, and domains as an electrostatic route for lowering depolarization energy.

### S1.3. After 2000: scanning probe localization of screening and surface reactivity

Kalinin and Bonnell used electrostatic-force and scanning-surface-potential microscopy to study local surface potential on $BaTiO_3$ and showed that the polarization bound charge can be essentially completely screened under ambient conditions [1]. The sign and magnitude of the observed surface potential were interpreted in terms of a double layer whose formation is strongly influenced by adsorbates. Subsequent Kalinin–Bonnell work connected local polarization to local reactivity and analyzed screening as a general limitation and physical mechanism in oxide-surface measurements [25,26].

In parallel, Giocondi and Rohrer demonstrated spatially selective photochemical reduction and separation of oxidation and reduction reactions on $BaTiO_3$ domains [27,28]. First-principles and surface-science work then made polarization-dependent adsorption quantitative. Kolpak, Grinberg, and Rappe predicted strong polarization control of the chemistry of $PbTiO_3$-supported Pt [29]; Yun and Altman measured polarization-dependent adsorption on $LiNbO_3$ [30]; Li and coworkers combined desorption measurements, surface-potential microscopy, and theory to demonstrate polarization-dependent physisorption energetics [31].

### S1.4. Synchrotron experiments and chemical boundary conditions

The converse direction was established particularly cleanly by in situ synchrotron experiments. Wang, Fong, Highland, Stephenson, and collaborators demonstrated reversible chemical switching of ultrathin $PbTiO_3$ by oxygen partial pressure [6]. Shin and coworkers showed that water exposure can change atomistic screening, surface reconstruction, and polarization in ultrathin $BaTiO_3$ [7]. Stephenson and Highland then formulated the thermodynamic theory in which ionic surface coverage is obtained from electrochemical equilibrium rather than imposed as a fixed charge [2]. Highland and coworkers mapped the equilibrium polarization and Curie temperature versus oxygen pressure and observed an intermediate-pressure region in which insufficient compensation stabilizes a low-polarization state [8].

### S1.5. Ferroionic states and later developments

The subsequent conceptual step was to treat ionic surface charge and ferroelectric polarization as coupled degrees of freedom. Mixed electrochemical–ferroelectric states were demonstrated and modeled in nanoscale ferroelectrics [3], while Morozovska, Eliseev, Morozovsky, and Kalinin introduced the explicit ferroionic-state description based on coupled LGD and Langmuir equations [4]. Kalinin, Kim, Fong, and Morozovska reviewed the broader surface-screening mechanisms and their consequences for polarization dynamics and domains [5]. Later work demonstrated chemically driven switching through polarization-selective surface bonding in $BiFeO_3$ [9], compared ionic and electronic screening regimes [10], and extended the surface-electrochemical control concept to $HfO_2$-based ferroelectrics [11,12] and aqueous ion adsorption on ferroelectric nanoparticles [13].

**Supplementary Table S1. Historical landmarks.**

| Period | Representative work | Principal result |
|---|---|---|
| 1949–1957 | Devonshire; Merz; Hooton–Merz; Känzig; Chynoweth; Landauer [17–19,40,49,51] | Bulk LGD thermodynamics, domain nucleation, polarization-selective etching, surface space charge, and electrostatic nucleation |
| 1962–1993 | Toyoda–Itakura; Cabrera et al.; Inoue et al. [20–24] | Atmosphere-dependent switching, domain-selective adsorption, and polarization-dependent catalysis/adsorption |
| 2001–2008 | Kalinin–Bonnell; Giocondi–Rohrer; Kolpak–Rappe; Li et al. [1,25–31] | Domain-resolved screening, reactivity, and quantitative polarization-dependent adsorption |
| 2009–2011 | Wang/Fong/Highland/Stephenson; Shin et al.; Stephenson–Highland [2,6–8] | Reversible chemical switching and electrochemical thermodynamic boundary conditions |
| 2017 onward | Yang et al.; Morozovska–Eliseev–Kalinin; Kalinin–Kim–Fong–Morozovska [3–5,9–13] | Mixed electrochemical states, explicit ferroionic thermodynamics, broader environmental control |

## S2. Model definitions, parameters, and numerical methods

### S2.1. Reduced model

The reduced calculations use $\alpha = 4(T/T_0 - 1)$, background dielectric parameter $\varepsilon_f = 0.25$, external dielectric parameter $\varepsilon_{\text{out}} = 0.10$, gradient coefficient $g = 0.005$, surface formation energy $\Delta = 0.80$, and maximum reduced surface charge $N_s = 0.35$. The second-order model uses $\beta = 1$, $\gamma = 0$; the first-order model uses $\beta = -1$, $\gamma = 1$. The surface has neutral, positive, and negative mutually exclusive states.

At chemical neutrality, the exact reduced chemical capacitance is

$$C_{\text{chem}}(T) = \frac{2N_s e^{-\Delta/T}}{T[1+2e^{-\Delta/T}]} \tag{S1}$$

The dense phase atlas in Figure 4(a)–(f) uses 181 temperature points and 241 chemical-pressure points for each transition order and thickness. Homogeneous stationary roots were compared by the appropriate grand potential. Finite-$k$ minimization used precomputed dense wave-number grids, and the weakly nonlinear stripe/checkerboard amplitudes were minimized analytically at each grid point. The phase colors therefore distinguish the lowest state within the restricted homogeneous/one-mode variational family; the overlaid line is independently the exact linear finite-$k$ spinodal within the columnar approximation.

### S2.2. $BaTiO_3$ parameters

**Supplementary Table S2. $BaTiO_3$ model parameters.**

| Parameter | Value | Source/status |
|---|---|---|
| $\alpha_1(T)$ | $4.124 \times 10^5$ [T − 388.15] J m $C^{-2}$ | Li–Cross–Chen [41] |
| $\alpha_{11}$ | $-2.097 \times 10^8$ J $m^5$ $C^{-4}$ | Li–Cross–Chen [41] |
| $\alpha_{111}$ | $1.294 \times 10^9$ J $m^9$ $C^{-6}$ | Li–Cross–Chen [41] |
| $\alpha_{1111}$ | $3.863 \times 10^{10}$ J $m^{13}$ $C^{-8}$ | Li–Cross–Chen [41] |
| $\varepsilon_b$ | 7.35 | phase-field parameterization [43] |
| $G_{phase}$ | $0.51 \times 10^{-11}$ J $m^3$ $C^{-2}$ | phase-field parameterization [43] |
| $G_{wall}$ | $8.88 \times 10^{-11}$ J $m^3$ $C^{-2}$ | sensitivity value calibrated to 7.5 mJ $m^{-2}$ at 300 K |
| effective $\Delta G_s$ | 0.20 eV | declared surface-chemistry benchmark |
| effective site area | 1 $nm^2$ | declared surface-chemistry benchmark |
| surface charge states | $z_+ = +2$, $z_- = -2$ | oxygen-related effective redox states |

The six film thicknesses are 1, 3, 10, 30, 100, and 300 nm. The full $BaTiO_3$ equilibrium maps use 181 temperature points between 260 and 440 K and 201 oxygen-pressure points over $-12 \leq \log_{10}(pO_2/\text{bar}) \leq 12$. Finite-k minimization uses a dense wave-number lookup spanning $10^5$–$10^{12}$ $m^{-1}$. The 300 K pressure cuts in Figure 7(a) and 7(c)were independently recomputed with 801 pressure points, 6001 polarization brackets, Brent refinement of every stationary root, and explicit grand-potential selection. The refined root selection removes the artificial small oscillations that coarse polarization interpolation can generate in thick-film chemical-capacitance cuts.

### S3. Analytical details of the domain-instability construction

For a finite-$k$ mode, the perturbation potential inside the ferroelectric is proportional to $\sinh(kz)/\sinh(kh)$, while the external potential decays as $e^{-k(z-h)}$. The linearized surface-charge response is $\delta\sigma = -C_{\text{chem}}\delta\phi_s$. Gauss' law then gives

$$\delta\phi_s = \frac{\delta P}{C_{\text{chem}} + k[\varepsilon_{\text{out}} + \varepsilon_f \coth(kh)]} \quad \text{(S2)}$$

which directly leads to Eq. (11) of the main text.

An annealed-then-frozen ionic boundary is distinct from the ion-poor limit. Let $\sigma_a$ be the mean surface charge equilibrated at the annealing condition. During a frozen measurement $\delta\sigma = 0$, so the frozen homogeneous polarization and finite-wave-vector kernel obey

$$f_L'(P_0^{\text{fr}}) + \frac{P_0^{\text{fr}} + \sigma_a}{\epsilon_f} = 0, \qquad \Lambda_{\text{fr}}(k) = f_L''(P_0^{\text{fr}}) + gk^2 + \frac{1}{hk[\epsilon_{\text{out}} + \epsilon_f \coth(kh)]} \quad \text{(S2a)}$$

The frozen-ion state therefore has zero incremental chemical capacitance but generally retains $\sigma_a \neq 0$ and the associated chemical bias. The ion-poor limit is the additional special case $\sigma_a \approx 0$. In the poor-screening large-$kh$ limit, minimization of Eq. (13) gives

$$gk_c^2 = \frac{1}{2h(\varepsilon_f + \varepsilon_{\text{out}})k_c} \quad \text{(S3)}$$

and the minimum positive electrostatic-plus-gradient contribution is

$$q_{\min} = 3gk_c^2 = \frac{3g^{1/3}}{2^{2/3}\left[h(\varepsilon_f+\varepsilon_{\text{out}})\right]^{2/3}} \tag{S4}$$

For a second-order material with $P_0 = 0$, the approximate critical thickness is therefore

$$h_c = \frac{3^{3/2}\sqrt{g}}{2(\varepsilon_f+\varepsilon_{\text{out}})[-\alpha(T)]^{3/2}} \tag{S5}$$

For the finite-$P_0$ stripe amplitude potential of Eq. (15), the local background polarization enters only through $\beta_{\text{eff}} = \beta + 10\gamma P_0^2$ at this order. Simultaneous stationarity and coexistence give Eq. (16). For the checkerboard ansatz, the corresponding moments produce Eq. (17). The pre-spinodal window exists only for $\beta_{\text{eff}} < 0$.

For the second-order $\phi^4$ wall, the exact profile and wall energy are

$$P(x) = P_s \tanh(x/\delta), \qquad \delta = \sqrt{2g/(-\alpha)} \tag{S6}$$

$$\gamma_{180} = \frac{2\sqrt{2}}{3}\frac{\sqrt{g}(-\alpha)^{3/2}}{\beta} \tag{S7}$$

The first-order $\phi^6$ wall can be reduced to a one-dimensional quadrature, and the $BaTiO_3$ eighth-order wall is evaluated analogously using the numerical bulk potential.

## S4. Numerical verification and nucleation calculations

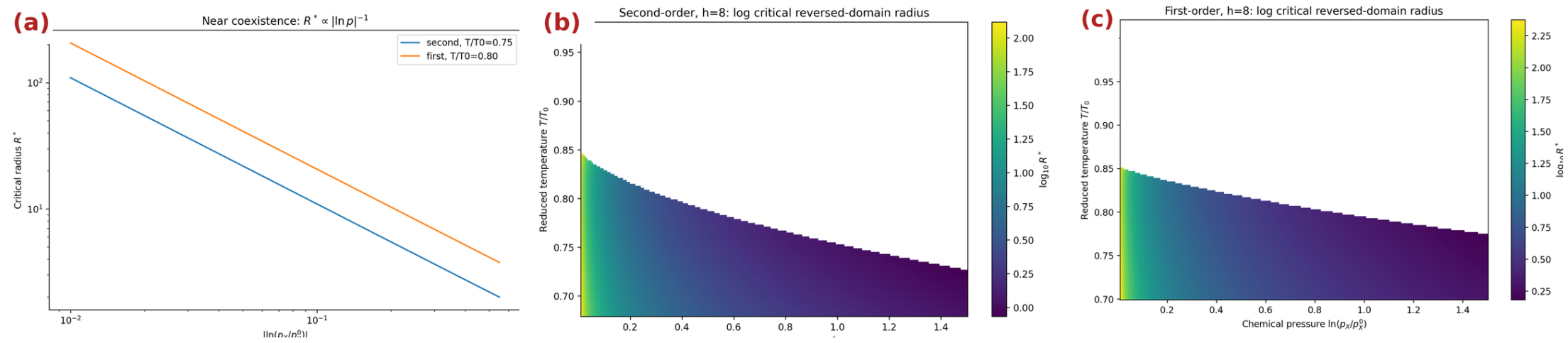

**Figure S1.** Static nucleation calculations retained as verification rather than as a central result. (a) Inverse chemical-bias scaling of the classical critical radius close to coexistence. (b),(c) Reduced second- and first-order critical-radius maps in regions where opposite homogeneous polar states coexist locally.

For a through-thickness cylindrical reversed nucleus with wall energy $\gamma_{\text{DW}}$ and grand-potential driving force per film area $\Delta\Omega_A$, the capillarity free energy is

$$\Delta G(R) = 2\pi R h \gamma_{\text{DW}} - \pi R^2 \Delta\Omega_A \tag{S8}$$

so that

$$R^* = \frac{\gamma_{\text{DW}} h}{\Delta\Omega_A}, \qquad \Delta G^* = \frac{\pi(\gamma_{\text{DW}} h)^2}{\Delta\Omega_A} \tag{S9}$$

Near symmetric chemical coexistence, $\Delta\Omega_A \propto |\ln(p/p_0)|$, giving the inverse-log-pressure divergence shown in Figure S1(a)–(c). These large classical barriers reproduce the logic of the Landauer nucleation paradox [49] and should not be interpreted as microscopic switching pathways in real defect-containing films.

## S5. Additional $BaTiO_3$ verification and profiles

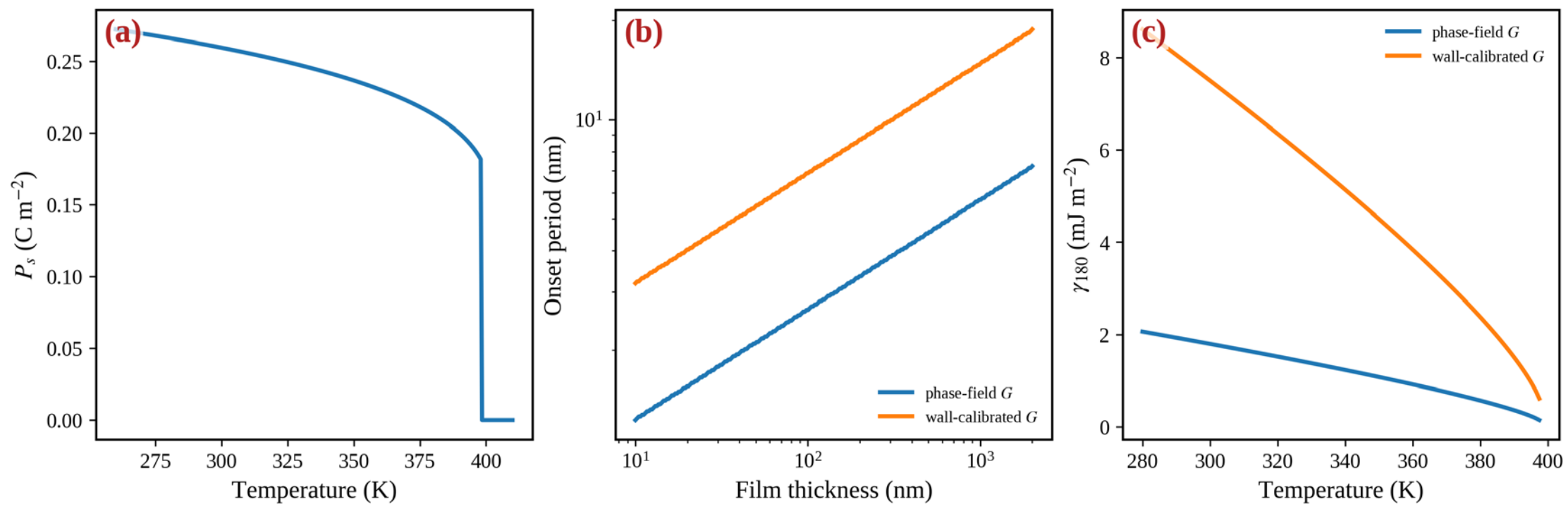

**Figure S2.** Additional $BaTiO_3$ dimensional checks. (a) Bulk spontaneous polarization from the eighth-order Li–Cross–Chen potential. (b) Finite-k onset wavelength in the ion-poor limit for the two gradient calibrations. (c) Calculated 180° wall-energy sensitivity versus temperature.

The $BaTiO_3$ bulk potential gives a cubic–tetragonal coexistence temperature of 398.15 K, a polarization discontinuity of 0.181 C m$^{-2}$, and a 300 K spontaneous polarization of 0.259 C m$^{-2}$. With the literature phase-field gradient coefficient the scalar wall energy at 300 K is approximately 1.8 mJ m$^{-2}$; the sensitivity coefficient $G_{\mathrm{wall}}$ is chosen to reproduce 7.5 mJ m$^{-2}$, within the lower first-principles range [44–46]. The wide spread in gradient/wall-energy estimates is the dominant uncertainty in converting the reduced domain-instability threshold to an absolute film thickness. Figure S2(a)–(c) collects these additional dimensional checks.

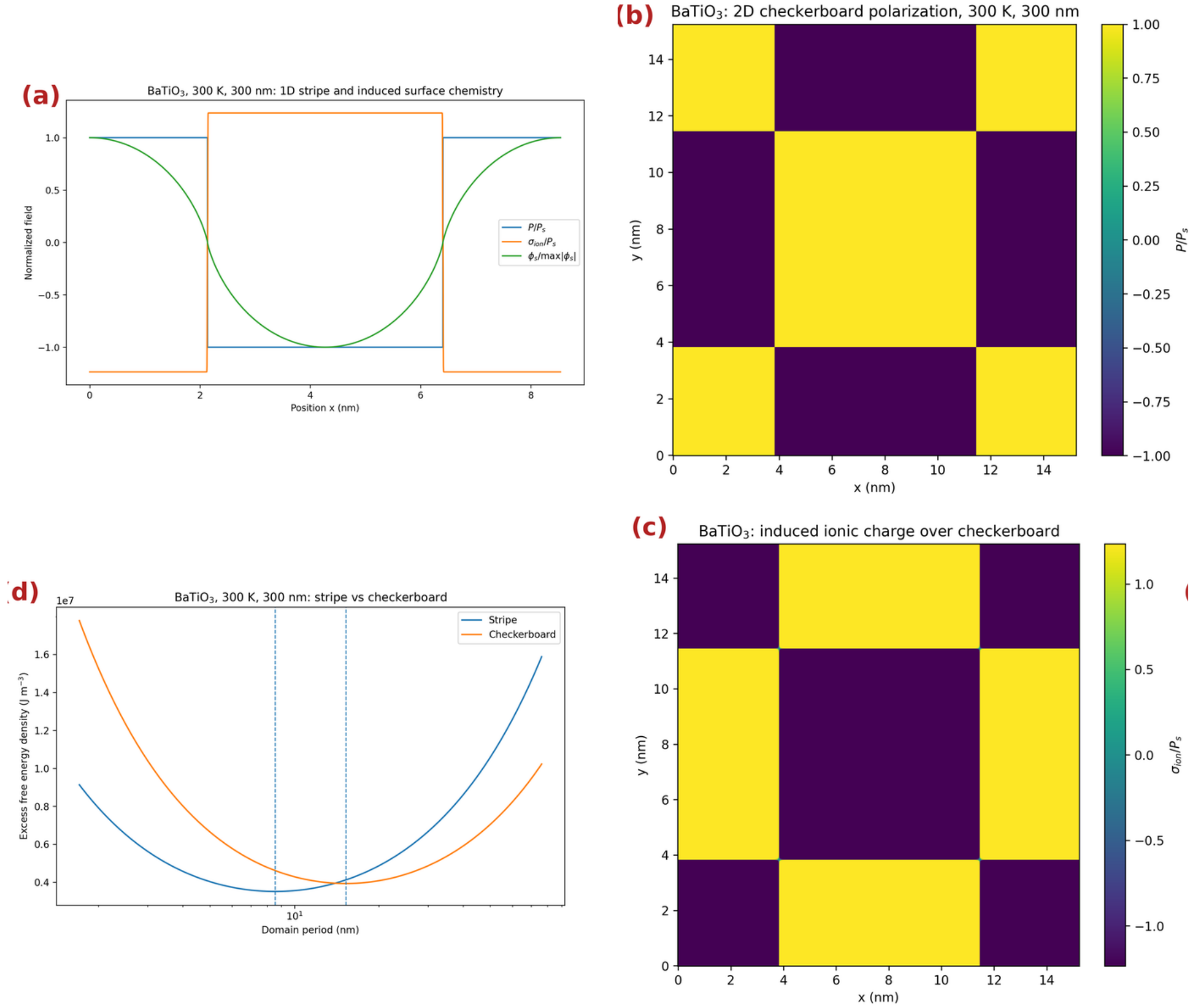

**Figure S3.** $BaTiO_3$ real-space and morphology checks at 300 K. (a) One-dimensional stripe polarization, potential, and surface charge. (b),(c) Two-dimensional checkerboard polarization and induced surface ionic charge. (d) Sharp-wall stripe/checkerboard free-energy comparison. These panels demonstrate the spatial electrochemical patterning but are not required for the main phase-diagram conclusions.

**Figure S3(a)–(d) shows the representative real-space morphologies and the stripe-versus-checkerboard free-energy comparison.**

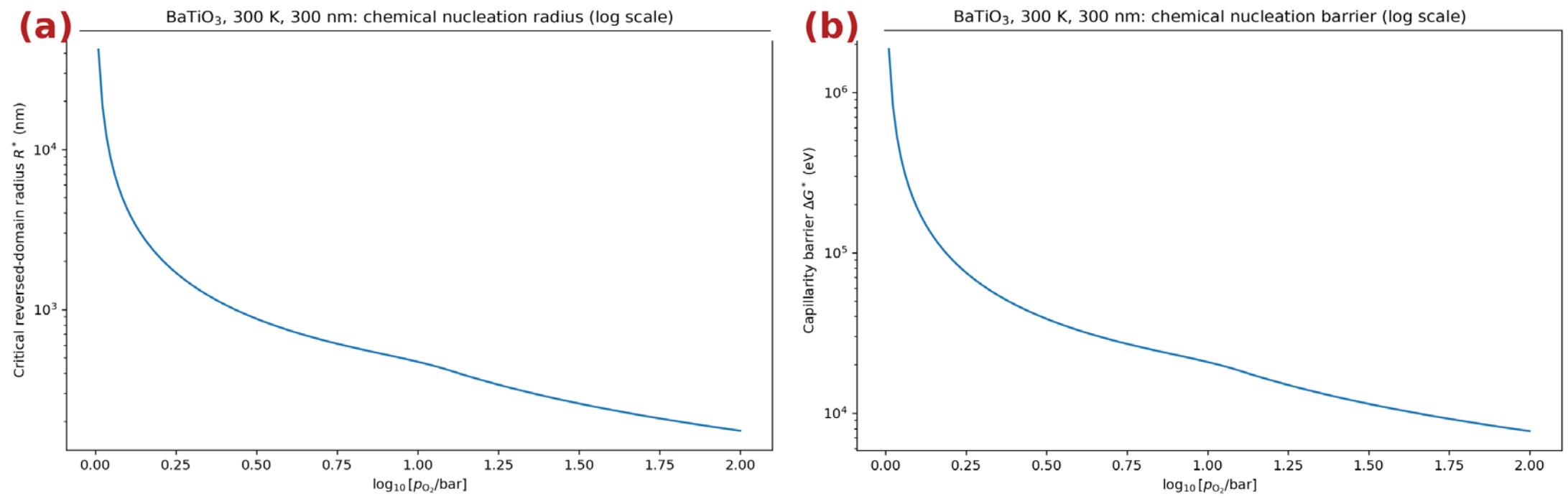

**Figure S4.** Classical chemical-bias nucleation in a 300-nm $BaTiO_3$ film at 300 K. (a) Critical radius. (b) Capillarity barrier. The very large values emphasize that homogeneous cylindrical nucleation is a limiting thermodynamic construction rather than a realistic microscopic switching mechanism.

**Figure S4(a) and S4(b) summarize the corresponding classical chemical-bias nucleation scales.**